# Nanoscale Surface Analysis on Second Generation Advanced High Strength Steel after Hot Dip Galvanizing

M. Arndt[1], J. Duchoslav[1], K. Preis[2], L. Samek[2], D. Stifter[1]

[1]*Zentrum für Oberflächen- und Nanoanalytik (ZONA) und Christian Doppler Labor für mikroskopische und spektroskopische Materialcharakterisierung (CDL-MS-MACH), Johannes Kepler Universität Linz, Altenbergerstraße 69, 4040 Linz, Austria*
[2]*voestalpine Stahl GmbH, voestalpine-Straße 3, 4031 Linz, Austria*

*corresponding author:* martin.arndt@jku.at

## Abstract

Second Generation advanced high strength steel is one promising material of choice for modern automotive structural parts because of its outstanding maximal elongation and tensile strength. Nonetheless there is still a lack of corrosion protection for this material due to the fact that cost efficient hot dip galvanizing can not be applied. The reason for the insufficient coatability with zinc is found in the segregation of manganese to the surface during annealing and the formation of manganese oxides prior coating. This work analyses the structure and chemical composition of the surface oxides on so called nano-TWIP (twinning induced plasticity) steel on the nanoscopic scale after hot dip galvanizing in a simulator with employed analytical methods comprising scanning Auger electron spectroscopy (SAES), energy dispersive x-ray spectroscopy (EDX), and focused ion beam (FIB) for cross section preparation. By the combination of these methods it was possible to obtain detailed chemical images serving a better understanding which processes exactly occur on the surface of this novel kind of steel and how to promote in the future for this material system galvanic protection.

## Keywords

## Introduction

Modern automotive industry has to fulfil with its materials two contradicting requirements by complying with the increasing demand of passenger safety and at the same time by reducing fuel consumption due to weight reduction of the vehicles. Both demands can be fulfilled by employing advanced high strength steels (AHSS) as material for structural body applications [1, 2]. Depending on their composition and deformation mechanism, these steels exhibit excellent mechanical properties and are often referred in literature as TRIP (transformation induced plasticity) or TWIP (twinning induced plasticity) steels [2-6].

The remarkable characteristics of the TWIP steel are based on the so called TWIP effect [7, 8], in which twinning planes are formed during tensile deformation. The microstructure of such steels is primarily constituted of austenite stabilized by high amounts of C and Mn. The tensile strength of these high manganese steels ranges from about 700 up to 1200 MPa with a total elongation from about 40% up to 110%. Not only the outstanding mechanical properties of this kind of steel render them very attractive but also the simple and cost effective formability by cold rolling instead of hot forming operations.

On the other hand TWIP steel is prone to the formation of manganese oxides on its surface during the annealing step in the production, which is needed for the adjustment of the material parameters. These oxides prohibit cost efficient hot dip galvanizing for the improvement of the corrosion resistance, because they hinder the growth of a continuous Zn layer on the steel surface. However, there are some approaches to improve the wettability of high manganese steel in a hot zinc bath: Blumenau *et al.* suggest a pre-oxidation step to create a FeMn mixed oxide on the surface of the steel. The Fe in the mixed oxide should then chemically be reduced during the annealing process and form a $Fe_2Al_5$ inhibition layer on top of the MnO, allowing the growth of the Zn layer. This approach was then successfully applied on simulator samples [9]. They also report an improvement of the wettability by changing the Al concentration in the bath, the bath temperature and the temperature of the steel [10]. Two other methods are suggested by Cho *et al.* [11], one with annealing at high dew points to provoke internal Mn oxidation and one with flash coating of the surface with pure Fe. Furthermore, Kavitha *et al.* [12] have observed that longer dwell times in the Zn bath can reduce the thickness of a MnO layer.

However, hot dip galvanised TWIP steel is so far not reasonably developed or let alone commercially available, with further need of material research and analysis to cope with this challenge. So far the reported analysis methods of choice were glow discharge optical emission spectroscopy (GDOES), conventional optical and scanning electron microscopy (SEM), transmission electron microscopy (TEM) and to a less extent X-ray photoelectron spectroscopy (XPS) [9-11, 13, 14]. Yet, up to now no high quality chemical mapping on the nanometre scale has been presented, which could aid in a detailed understanding of the chemical and structural surface formation. Therefore, the focus of this work is put on new insights in this topic with additional and complementary characterization methods and a combined analytical effort using focused ion beam (FIB) preparation, energy dispersive X-ray spectroscopy (EDX) and, especially, high-resolution scanning Auger electron microscopy (SAM) and spectroscopy (SAES).

# Experimental

## *Sample material*

The sample material used in this study is a second generation advanced high strength steel (AHSS) directly taken from industrial production. In detail, the material is a nano-sized twinning induced plasticity (nano-TWIP) steel with a bulk concentration of more than 83,106 wt.% Fe, 0.79 wt.% C, 15.8 wt.% Mn, 0.05 wt.% Si, <0.05 wt.% Al, 0.03 wt.% P, 0.002 wt.% Ti, 0.022 wt.% Nb, 0.035 wt.% N and 0.03 wt.% Cr. Furthermore minor components include 0.025 wt.% Ni, 0.003 wt.% V, 0.0025 wt.% S, 0.016 wt.% Cu, 0.017 wt.% Mo, 0.005 wt.% Sn, 0.001 wt.% Zr, 0.005 wt.% As, 0.0001 wt.% B, 0.009 wt.% Co, 0.001 wt.% Sb and 0.0001 wt.% Ca. The sample material was annealed in a 100% $H_2$ atmosphere at a dew point of -45°C in a galvanizing simulator and immersed in a hot zinc bath. The zinc bath contains small amounts of Al in the order of 0.2 wt.% to aid the formation of an inhibition layer of $Fe_2Al_5$ between the steel substrate and the Zn overlayer.

From the galvanized steel sheet sample pieces with the dimension of 8 x 8mm$^2$ were cut out and cleaned in an ultrasonic bath to remove adventitious carbon based surface contaminations, which originate from sample handling in the production line. These contaminations especially accumulate in trenches and craters on the surface and make Auger spectroscopy in these areas impossible, as our experience has shown [15]. In detail, the beakers used for cleaning in the ultrasonic bath are at first washed with a mixture of 100 ml sulphuric acid (Merck, 95-97 %, p.A.) and 100 ml hydrogen peroxide (Merck, 30 %, stabilized for synthesis). For the following cleaning steps, the samples are consecutively placed for 15 minutes in organic solvents, namely tetrahydrofuran (Sigma-Aldrich, anhydrous, ≥ 99.9 %, inhibitor free), isopropanol (VWR, 99.9 %) and ethanol (VWR, 99.9 %), with each step carried out twice and by always using a fresh organic solvent.

### *Analytical methods*

The main analytical work in this study was performed with a scanning Auger electron microscope JAMP 9500F (from JEOL). The system operates in a similar way as a standard hot field emitter scanning electron microscope (FESEM) with three mayor exceptions. Firstly, the pressure in the analysis chamber is in the ultra high vacuum region below $10^{-7}$ Pa. Secondly, there is a hemispherical electron energy analyser to perform spectroscopic Auger electron measurements in the range from 0 to 2500 eV with a spatial resolution down to 8 nm. As to the detector itself, the AES spectra are simultaneously recorded by 7 channeltrons in parallel. Finally, there an argon sputter gun with ion energies between 0.5 and 3 keV can be employed for sputter depth profiling. Alternatively, this gun can be used for charge neutralising by using ion energies between 10 and 50 eV. The AES spectra as well as the image data were analysed with software provided by the system manufacturer (Spectra Investigator and Image Investigator, JEOL).

The second analytical system used in this study is a 1540 XB from Zeiss, being a FESEM combined with a gallium focused ion beam (FIB) source. This device was used for cutting lamellas in the dimension 10 μm x10 μm x 1 μm out of the surface in order to be analysed in a second step by AES. The microscope was also used for overview images from these lamellas due to the high lateral resolution of the probe going down to 1 nm and the in-lens detector keeping this high resolution in the image by collecting only secondary electrons created by the primary electron beam and not from the backscattered electrons (SE1). The SEM is also equipped with a SiLi energy dispersive X-ray (EDX) detector from Oxford to measure the characteristic X-rays from the sample surface.

# Analytical results

### *General overview via EDX*

Due to the reduced wettability of the TWIP material, as indicated above, the surface is not fully covered with zinc. In general, there are surface regions exhibiting Zn droplets extending some millimetres in diameter, while other areas appear to be fully uncovered. Figure 1 shows such an uncovered area between two Zn droplets on microscopic scale. For the EDX analysis the accelerating voltage of the incident electron beam was set to15 kV in order to excite the

$K_\alpha$ lines of Mn, Fe, and Zn. It is also worth mentioning that at this accelerating voltage the information of the X-ray signal originates from a maximum sample depth of 1 μm.

In detail, the maps show the distribution of all measured elements from the EDX sum spectrum. The Zn signal is mainly visible on the two large Zn droplets at the left-hand and right-hand side of the image, interrupted by smaller traces from the area in between. This proves that the bulk of the Zn droplets consists of pure metallic Zn: especially, within the left Zn droplet there is no other element visible. However, on the right Zn droplet there is additional to the Zn signal also a mixture of Al und O detectable, leading to the assumption that aluminium oxide is covering the surface there with a dimension of more than the typical aluminium oxide thickness on Zn of 5 nm. It is shown later by Auger mappings (Figure 4) that the aluminium oxide lies on the surface and is not covered by Zn. The element distribution in the region between the Zn droplets shows a homogeneous distribution of Mn, O, and Fe, whereas the Al signal is especially located at the border regions of the Zn droplets. Particularly, the appearance of Fe gives a hint that this region is the uncoated original surface of the TWIP steel.

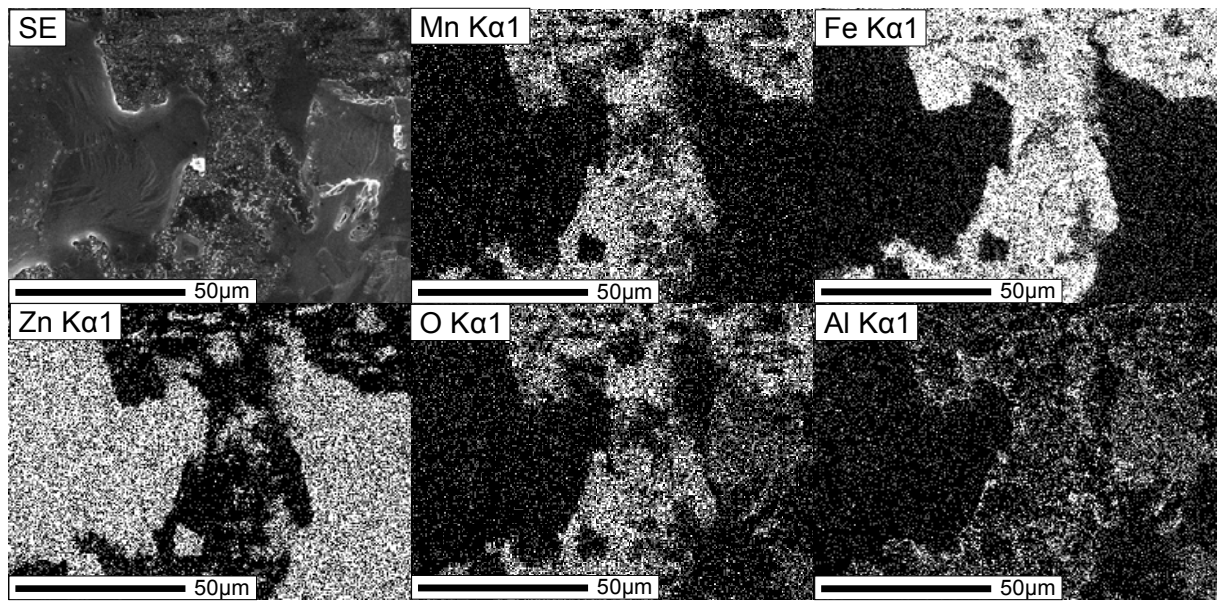

**Fig. 1:** SE image and EDX mappings of a surface area between two Zn droplets on the left-hand and right-hand side. The accelerating voltage of the primary electron beam is 15 kV. Size of EDX mappings: 256x192 pixels.

In Figure 2 a magnification of the region from the middle of Figure 1 is now shown. In order to achieve a higher lateral resolution the acceleration voltage was decreased to 5 kV, which leads to two consequences which need to be discussed. Firstly, the energy is too low to excite the $K_\alpha$ lines of Mn, Fe, and Zn. Therefore, in this mapping the $L_\alpha$ lines were used. It should also be mentioned that the Fe $L_\alpha$, Mn $L_\alpha$ and O $K_\alpha$ lines cannot be completely separated because of their overlapping energies. Secondly, the obtained depth information is not the same as in Figure 1: the penetration of the primary beam is for 5 kV around 100 nm instead of 1 μm, so these mappings are more surface sensitive than those in Figure 1.

Furthermore, in the secondary electron image crystalline structures are visible. Most of these structures consist of pure Zn, whereas some crystals are formed by Al. In addition, the distribution of Fe is no longer as homogeneous as in Figure 1 but shows coexistence with Al. However, similar to Figure 1, there is also here a homogeneous distribution of O and Mn present.

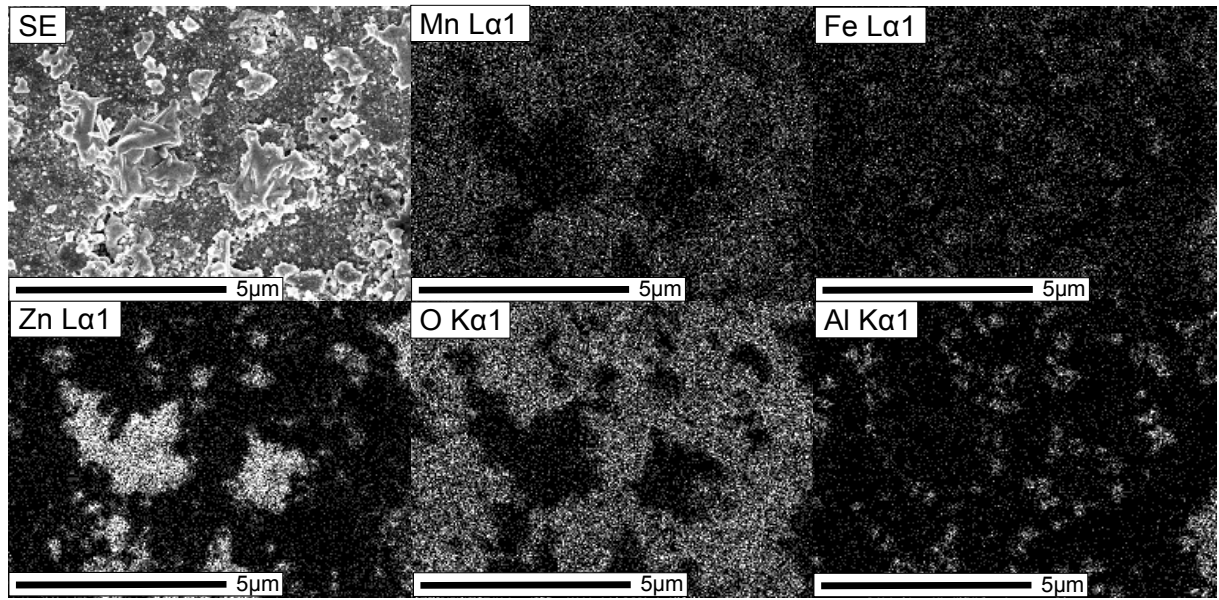

**Fig. 2:** SE image and EDX mappings of the region between the large Zn droplets shown in Figure 1. The acceleration voltage of the primary electron beam is 5 kV.

## *AES surface spectra*

For a more detailed and complementary analysis, the same surface area which was analysed by EDX was also investigated by AES. The parameters for the primary electron beam were set to 30 kV and 10 nA for the spectroscopic measurements in order to also achieve a high lateral resolution for the subsequently performed surface mappings without the need of changing the beam conditions. The surface scans were made directly after the ultrasonic cleaning treatment without argon sputtering. By this method only carbon contaminations were removed from the surface but the topmost oxides remain unchanged. The spectra were in each case taken out of an area as marked in the corresponding image with a rectangle exhibiting the same colour as the spectrum on the right-hand side.

From the upper image on the left it can be seen that the red spectrum shows the typical characteristics of a surface of a hot dip galvanised steel, consisting mainly of a homogeneous $Al_2O_3$ layer with an approximate thickness of 5 nm. This layer thickness is an estimation because the spectra look similar to AES spectra taken from industrial produced hot dip galvanised steel samples, where the layer thickness was calculated to be 5 nm using the inelastic mean free path of Zn $L_3M_{45}M_{45}$ electrons in $Al_2O_3$ and the Beer-Lambert law (results not shown here). Below this aluminium oxide layer metallic bulk Zn can be found, as derived from the spectrum considering two effects. At first, the Zn peak in the upper red spectrum - at an energy of 990 eV - is not as good developed as the ones in the other spectra, especially when compared to the Zn signal of the red spectrum from the plot in the bottom of Figure 3. At the same time, a broad hump is visible at lower energies between 700 and 950 eV, proving that a large reservoir of Zn lies below the surface in which Zn Auger electrons are generated and inelastically scattered when passing through the overlying aluminium oxide layer. The second effect is the clear appearance of the Al-LVV signal in the energy region below 100 eV. These electrons have an extreme low inelastic mean free path in solid material and are therefore only visible when they are originating from the topmost surface.

The green spectrum in the upper plot is rather similar to the red one: there is also an $Al_2O_3$ layer on top of metallic Zn. However, one difference is the additional appearance of Mg in the spectrum. The shape of the Mg peak also shows that Mg is located on the surface most probably in the form of MgO, as shown in one of our previous works [15]. Mg might in this case come from minor contaminations of the Zn bath in the simulator, since no Mg was nominally added to the steel. Due to its affinity to oxygen, Mg is most probably located on the topmost surface. So it is not detectable by EDX, but only by AES.

The blue area in the upper image shows a slightly different spectrum than the two other ones: not only is the Zn signal is much clearer developed and does not exhibit the broad hump in the lower energy range of the Zn region as the two earlier mentioned spectra, hinting at the fact that the Zn signal originates in the blue curve more from the surface, but additionally Mn (besides traces of sulphur) is appearing in the spectrum.

The lower picture now shows a magnification of the area between the two larger Zn droplets. The three spectra taken from this area are quite similar to the blue one from the upper image with the Zn signal being clearly developed and with missing signs of inelastic scattering from an overlayer. Furthermore, Al, Mn, Mg, and O appear in all spectra and there is surprisingly not much difference between the red spectrum taken from a crystal and the other ones taken from the darker regions.

Also further spectra were taken from other areas of the sample surface, which are not shown here, with the remarkable finding that - opposed to what EDX analysis suggests - in none of these spectra Fe could be detected.

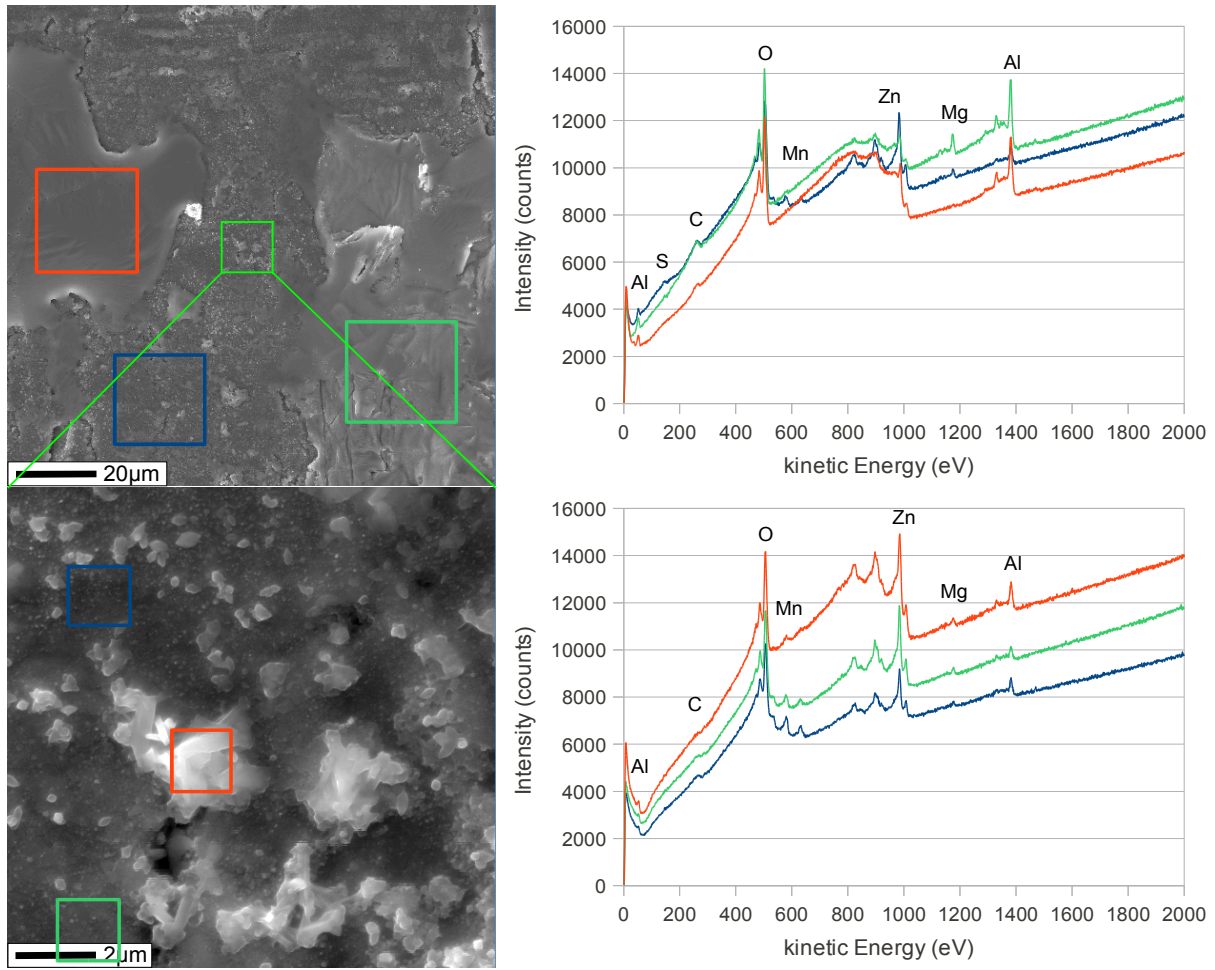

**Fig. 3:** SE images and corresponding AES surface spectra from the same regions as analysed in Figure 1 and 2. The lower area is a magnification of the middle of the upper one. The accelerating voltage of the primary beam is 30 kV. Al and O is present in all Auger electron spectra from the surface, whereas no Fe is detectable.

## *AES surface mapping*

After determining the elements in the first nanometres on the surface AES mapping was performed to obtain their lateral distribution. The result is shown in Figure 4, with Zn, O, and Al found nearly everywhere. Furthermore, on top of the Zn droplets there is always an $Al_2O_3$ layer but with varying thickness. In the upper right corner of the scanned area the layer is even so thick, that nearly no Zn is detectable, which is in good agreement to the EDX-results as presented in Figure 1, where this is the only area on top of a Zn droplet in which Al and O could be found. A homogeneous Mg distribution within this area is also worth mentioning. Mn is present only in the valley between the Zn droplets. Finally, C and S are found at the edges of the Zn droplets. It seems that these are residues of the carbon contamination that was removed by ultrasonic cleaning. These elements could not be detected by EDX.

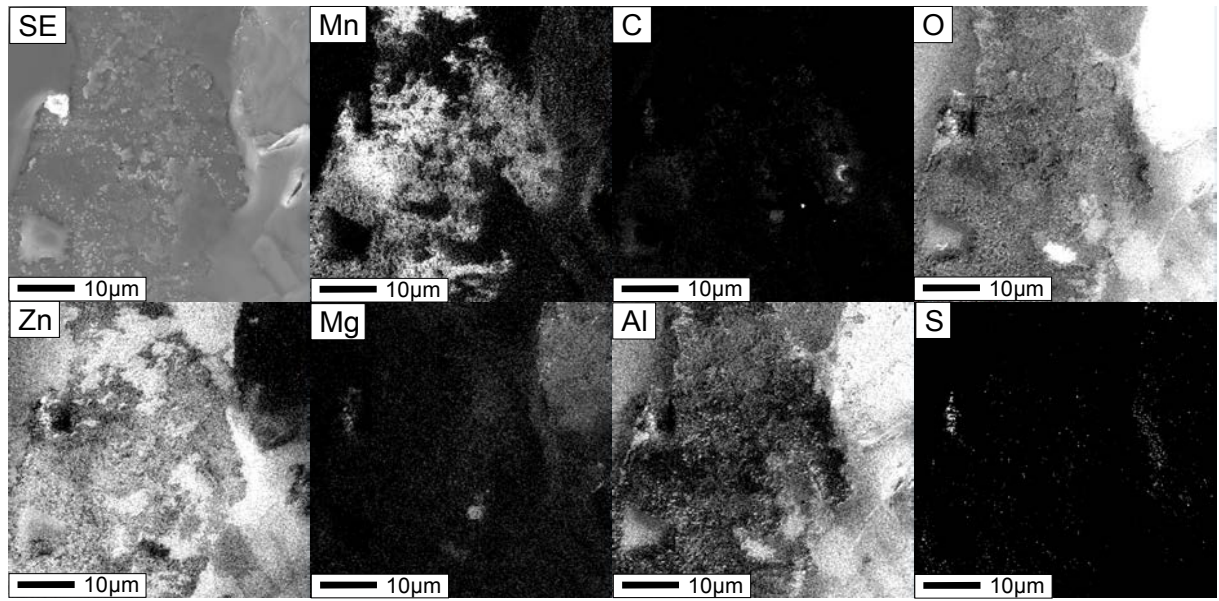

**Fig. 4:** SE image and corresponding AES surface mappings of all elements found in the survey spectra shown in Figure 3. C and S are residues of the carbon contamination and found on the steep edges of the Zn droplets.

## *Cross section preparation via FIB*

For a further analysis why some areas exhibit coverage by Zn whereas others not as well as for a correct interpreting of the previous results it is necessary to have a view into the depth of the material. This was accomplished by focused ion beam (FIB) preparation, as seen in Figure 5. However, the analysis of the steel surface in an area under a Zn droplet and a direct comparison with a depth analysis from an uncovered region surface is not feasible due to the extended thickness of more than 20 µm of these droplets, which is too thick for cutting with the Ga ion beam. Therefore, one investigated region is located at the edge of a Zn droplet (Figure 5a) and the second one from the area with the small Zn containing crystals (Figure 5c) which was analysed before by EDX (see Figure 2) and AES (Figure 3, lower part).

In order to avoid curtaining effects the surface was at first covered by a platinum carbon layer deposited from gas phase via the Ga beam. Furthermore, it is also necessary to take out lamellas from the surface and not just to analyse the direct cut in the surface because of severe preparation artefacts within the Zn material: distinctive holes are appearing in the Zn droplets, as seen in Figure 5a, which is not the case for the preparation of lamellas. Usually, the whole procedure of the lamella preparation is needed for further TEM investigations, which also calls for thinning of these lamellas down below 100 nm. Fortunately, the lamella thickness plays no role for AES investigations on the cross-section, since the combination of the different materials would make such thinning processes rather challenging.

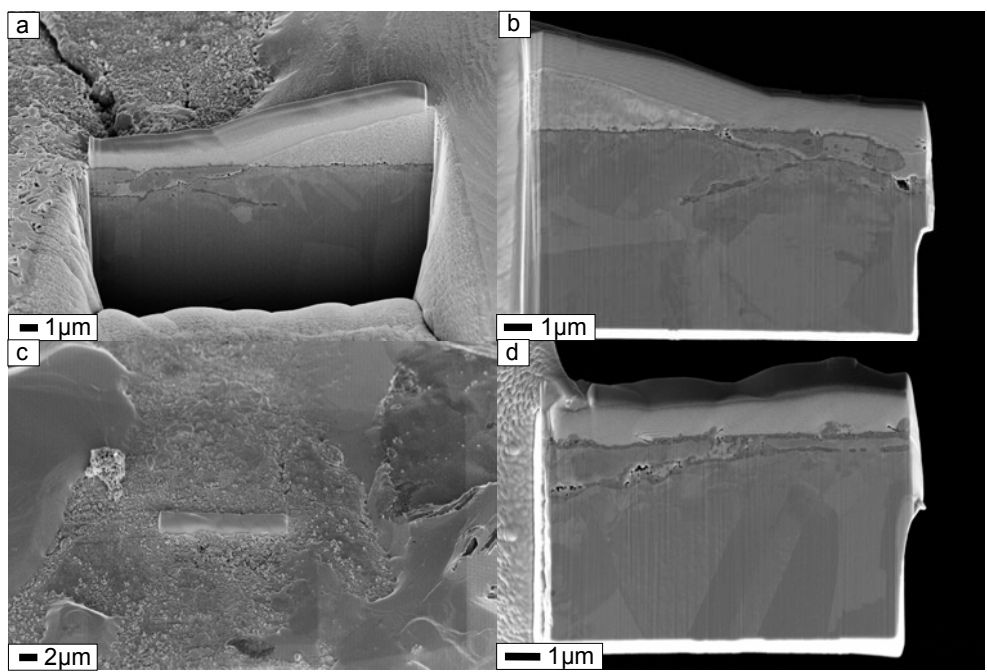

**Fig. 5:** Lamellas (b) and (d) were cut out from two different areas (a) and (c) by focused ion beam. The lamellas are ion polished and shown from the backside (mirrored by the y-axis).

## *AES mapping of the FIB lamellas*

Before the results of Auger analysis of the FIB lamellas will be explained in detail some general remarks concerning the measurements need to be made. Figure 6 exemplarily shows the results of a part of the FIB lamella from Figure 5b. As one can see, the secondary electron image is low in contrast. Despite this, different structures can be distinguished (represented by grey shades in the image) which can be separated by Auger mapping. Most of the mappings were made by using the most intensive Auger lines of each element: O KLL, C KLL, Pt MNN, Zn LMM, Fe LMM, and Al KLL. Only for Mn the $L_3M_{23}M_{45}$ line at 582 eV was used instead of the usually taken Mn $L_3M_{45}M_{45}$ line at 631 eV due to a complete overlap with signals from Fe and O. Furthermore, all mappings were made in constant analyser energy CAE mode with constant pass energy. In all cases the 7 available channeltrons of the detector were simultaneously used and the low energy part of them was defined as the peak signal P whereas the high energy part was defined as the background signal B. The shown intensity I in the mappings is then defined as I = (P-B) / (P+B) in order to remove an influence of the topography on the signal.

In order to reduce the number of mappings for interpretation, an adapted kind of presentation style was applied for images combining several elements or compounds (*e.g.* Figure 6, bottom, right). Usually in the image investigator program, which is provided by JEOL for the analysis of Auger mappings, three elements are combined in one image by using the red (R), green (G), blue (B) base colours. In order to show that Al is only present where Fe is found, the RGB combination of Mn, Al, and Fe was applied. To this combination Zn was added as yellow colour by using the GNU Image Manipulation Program (GIMP). In this way it is still clearly visible that Al is always appearing together with Fe, since there is only one green tone present. The C and Pt mappings are not included because these elements are only originating from the lamella preparation.

In addition, it has to be stated that oxygen is a rather critical element in these measurements and also not shown in the combined image, since its concentration is not stable: one can see in the mappings that all of the Mn is present where also a high O concentration is found, so that one can speak of a manganese oxide phase. However, oxygen can also be found in a lower concentration where the Zn resides. Furthermore, the concentration of oxygen is growing with increasing measuring time although the vacuum in the main chamber is below $10^{-7}$ Pa. For an estimation, the measurement time for one image with a resolution of 256x256 pixels and a dwell time of 10 ms per pixel is approximately 15 minutes including time for drift correction. The total measurement time for 7 elements with 10 repetitions sums up to approximately 18 hours. At the beginning of the measurement no O is visible in the Zn regions, but after some repetitions the highest oxygen concentrations can be found there. Besides, from EDX measurements it is known that the O concentration in the Zn rich regions is low (cf. Figures 1 and 2). Therefore, in the following measurements it is assumed that Mn is in reality an MnO phase and the O signal has been omitted.

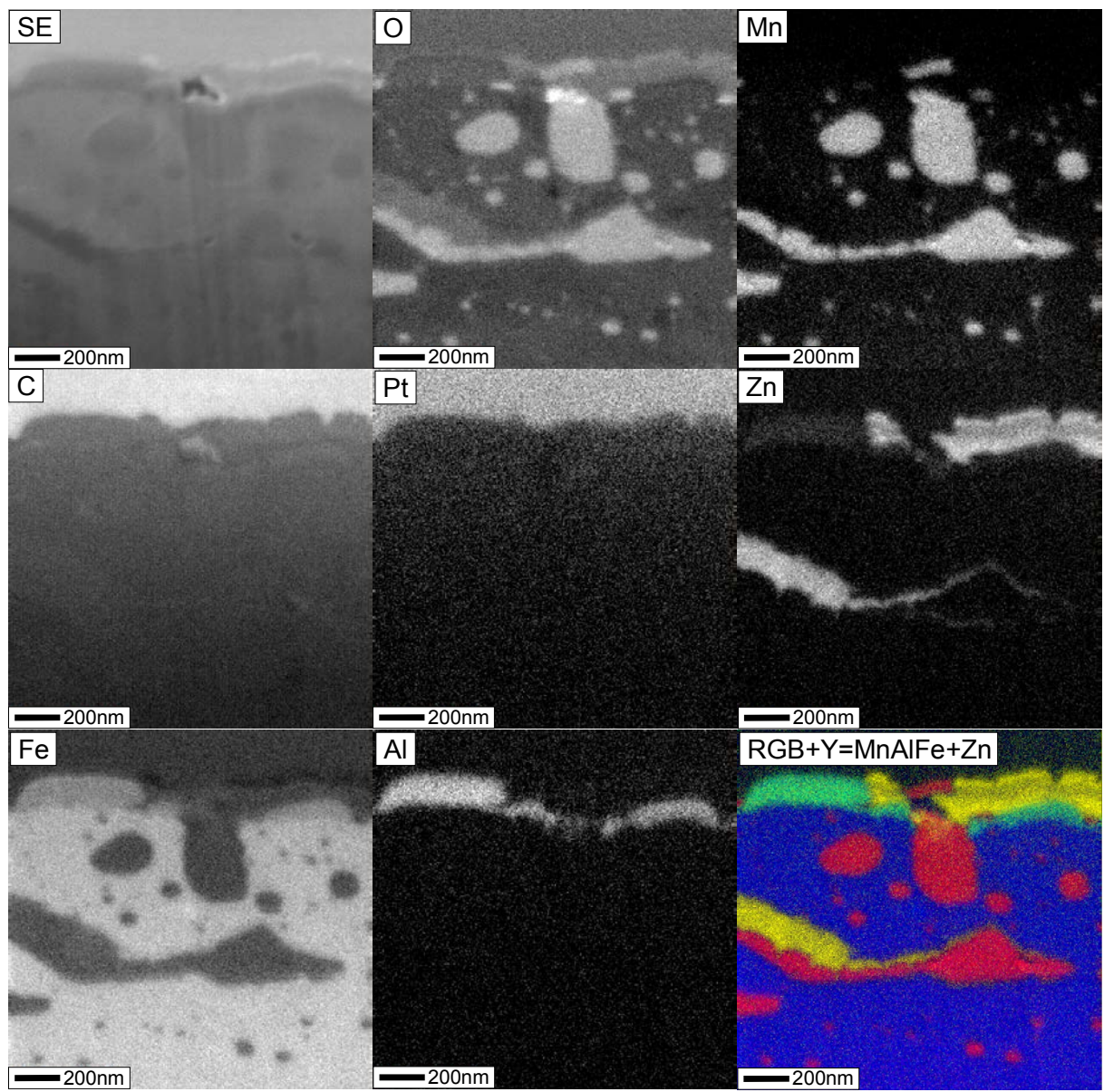

**Fig. 6:** SE and Auger electron images of a part of the FIB lamella, shown in Figure 5b. The most important elements, namely Mn, Al, and Fe, were combined in a single image (lower right corner) by giving each element a base colour: red (R) to Mn, green (G) to Al, and blue (B) to Fe. In addition to this standard RGB-presentation, the Zn data was added as yellow colour.

## *AES results of the FIB lamella from the Zn droplet free area*

On the lamella presented in Figure 5d several AES mappings were performed, as depicted in Figure 7 and giving an overview of the surface region of the steel (Zn droplet free region). Evidently, a continuous MnO layer is formed on top of the steel substrate with a film thickness between 100 and 200 nm. In addition, it is also visible that MnO layers are even situated below the surface. In these layers some holes can be observed (cf. Figure 5a) as well as surprisingly also Zn (yellow). On the surface itself small dots (yellow and cyan) are present.

Figure 7c and 7d give magnified areas from the surface region. Figure 7c is a cross section of the left crystal shown in Figure 2 and in the lower image of Figure 3. This crystal consists of two phases, namely pure Zn and a mixture of Fe and Al. From literature the latter phase is known to be $Fe_2Al_5$ forming the so-called inhibition layer [9, 10]. Figure 7d now shows in detail that Zn can be found below the surface. The simultaneous presence of inhibition layer crystals in the depth, although the steel is nominally free from Al, leads to the conclusion that here a void was filled with liquid Zn from the bath. Also mentionable is the fact that the Zn can grow directly neighboured to or embedded within the MnO.

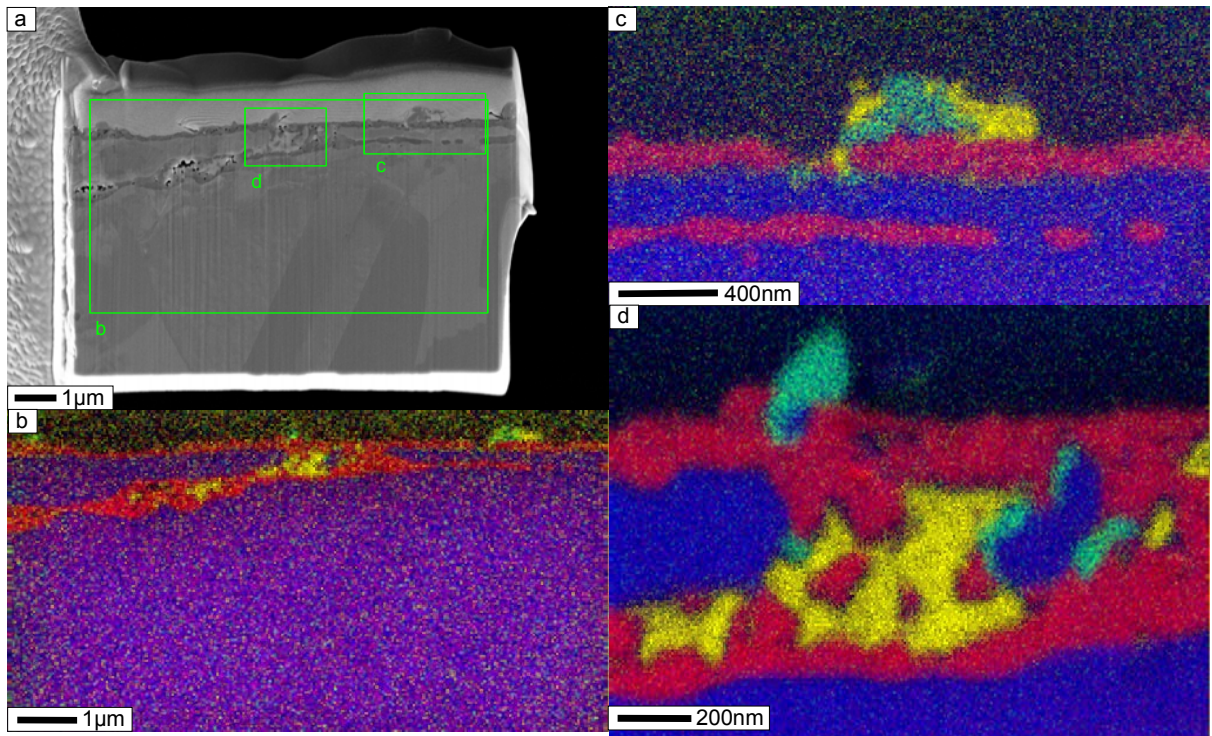

**Fig. 7:** SE overview image and AES mappings of three different regions of a FIB lamella from a Zn droplet free region. The colours refer to red MnO, cyan $Fe_2Al_5$, blue Fe, and yellow Zn. A homogeneous MnO layer on the top of the steel with a thickness between 100 and 200 nm has been formed and Zn is present below the surface.

## *AES results of the FIB lamella from the Zn droplet area*

The AES mappings from the FIB lamella taken from the edge of a Zn droplet, as presented in Figure 8, give a mixture of the $Fe_2Al_5$ phase (cyan) and the MnO phase (red) on top of the steel. Especially below the Zn droplet the existence of an inhibition layer is essential. However, at the edge of the Zn droplet already a MnO layer, similar as depicted in Figure 7, appears with voids [9] between MnO and the metallic Zn, as seen in Figure 8c. Figure 8d shows an area of a steel platelet where no MnO layer was formed. Instead MnO is building round inclusions inside Fe. On top of the platelet the cyan-coloured inhibition layer could develop and consequently also Zn is found there. On the other hand, a MnO layer is formed below the steel platelet which is still in contact with bulk steel material. This observation is confirmed by observing that the Zn was filling the former hole below the steel platelet.

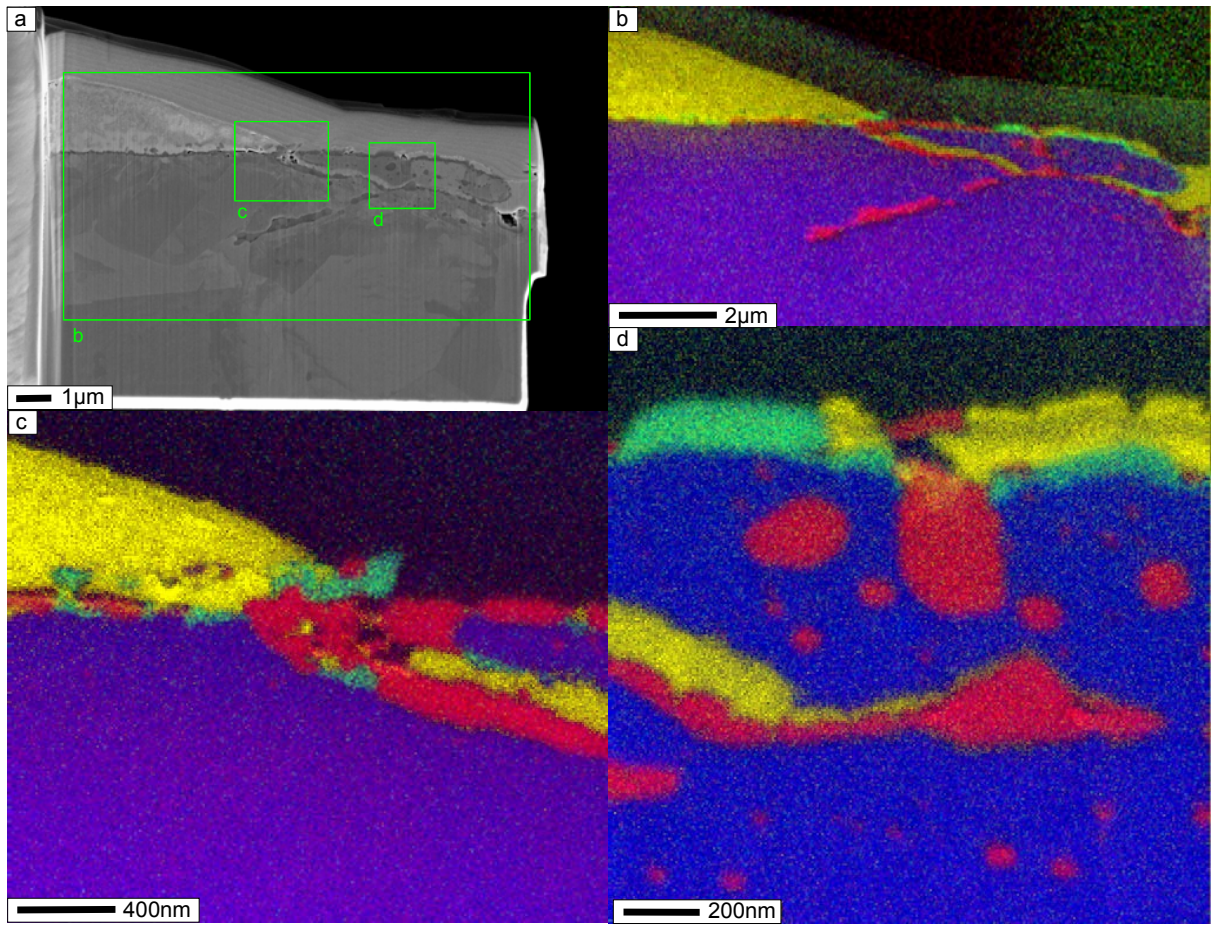

**Fig. 8:** SE overview image and AES mappings of three different regions of a FIB lamella from a region at the edge of a Zn droplet. The colours represent: red MnO, cyan $Fe_2Al_5$, blue Fe, and yellow Zn. A mixture between the $Fe_2Al_5$ inhibition layer and the MnO layer on top of the steel is visible.

### *AES depth profiling on a Zn crystal*

The shape of some Zn crystals found on the surface does not conform to those reported in literature [9, 13]. It is expected that pure Zn forms droplets instead of distinctively shaped crystals. To clarify this point, an AES sputter depth profile was performed on such a Zn crystal, as depicted in Figure 9. It has to be noted that a depth profile with Auger from an area which is in the order of magnitude of the electron beam diameter can be taken, although the diameter of the sputter crater itself was approximately 1 mm in diameter. For the depth profiling experiment a multi spectrum with different energy ranges was measured and subsequently argon ion sputtering was carried out. This procedure was repeated 19 times with the result shown in Figure 9b. The sputter parameters were 1 kV argon energy with 1 minute sputtering time per cycle. The sputtering speed was calibrated by means of a 100 nm thermal $SiO_2$ on a Si wafer to be approximately 4 nm per minute. The measured intensities from the spectra of Figure 9b were then converted to concentrations by dividing the values by the respective relative sensitivity factors (RSF) and by normalising the summed concentrations to 100%.

From the diagram of Figure 9c it is obvious that the Al and O signal are coming only from the topmost surface and disappear with increasing depth. The same finding holds true for C. On the other hand the Zn signal increases with depth. However, the most important observation is that the Fe signal begins to appear below the surface, which means that the investigated Zn structure is in fact a ZnFe mixed crystal. The Zn Fe ratio can not be taken directly from the depth profile data because it is known that Zn is most probably preferentially sputtered so that the real Fe concentration is in fact lower than the measured one, as demonstrated before for Mg in Zn [15]. With this knowledge and from comparisons with literature these crystals are now attributed to the $FeZn_{13}$ ζ phase [9, 13, 14]. Additional sputter depth profiles were also performed on other areas (results not shown here) and especially on sputtered Zn droplets no Fe could be found.

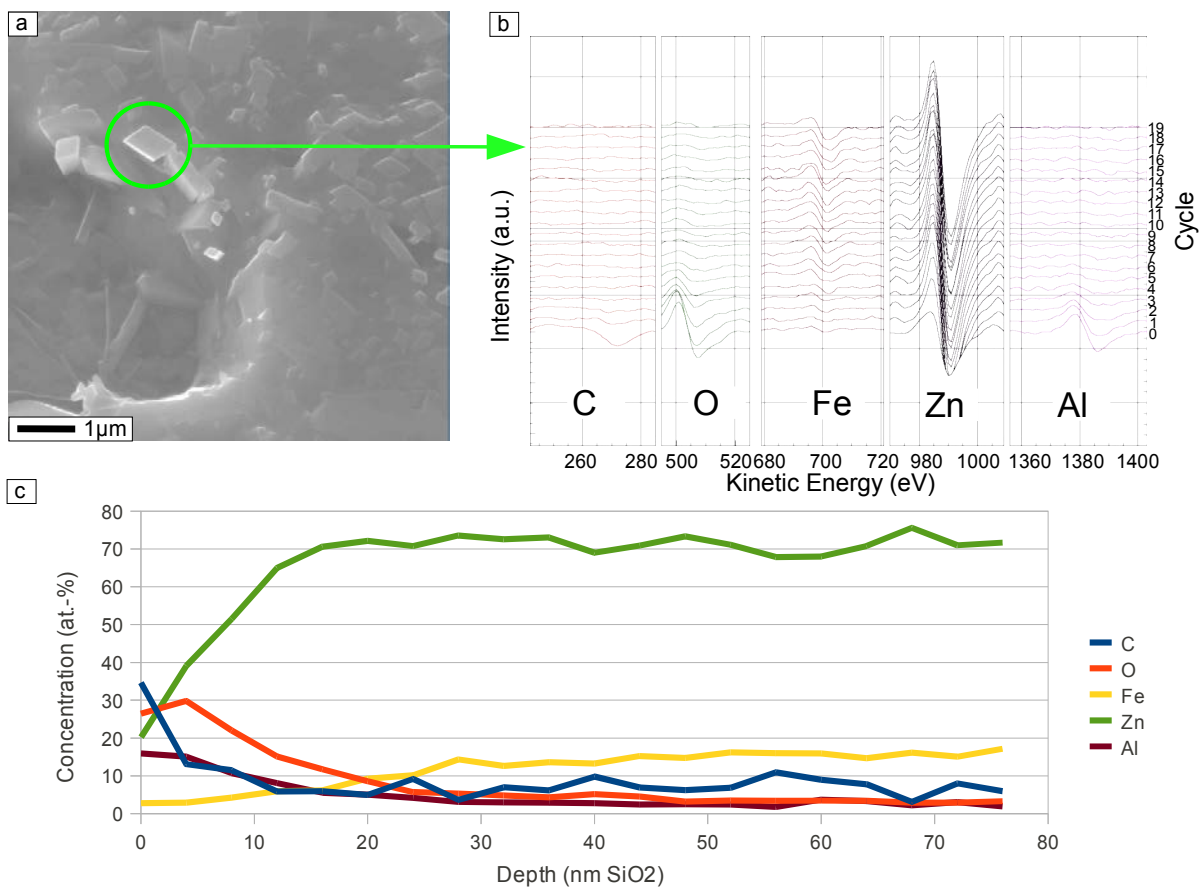

**Fig. 9:** The derivative AES spectra of 19 sputter cycles of a single Zn crystal (a) are shown in (b), with the calculated sputter depth profile of these data shown in (c). $Al_2O_3$ and C are only found on the surface while in the depth a mixture of Fe and Zn appears, attributed to the ζ phase.

## *AES mappings on sputter cleaned surface*

After the removal of the outermost aluminium oxide layer by argon ion sputtering the surroundings of the Zn crystal was also explored by high quality Auger mapping. In the images the number of pixels in each direction is 512 and each image acquisition was repeated 10 times in order to maximise the signal to noise ratio by averaging, with the result shown in Figure 10.

The AES mappings of Zn and Fe demonstrate that the signal is sufficiently clean in order to distinguish between $FeZn_{13}$ and pure Zn in the crystalline area in the lower left part of the image. Furthermore, the Al signal is not distributed over the whole area but shows sharply separated regions with high concentrations which correspond to regions exhibiting also a high Fe concentration. Mn is found outside these areas with augmented Fe, Zn, and Al concentrations. A puzzling observation can be made for the O signal: the highest amounts are neither found in the Mn rich areas nor at some round bright structures visible in the secondary electron image which could be oxides. Instead, the highest oxygen signal is located within the AlFe phases. This is, as already indicated above, a measuring artefact due to a reaction of small amounts of O from the residual gas with the surface after a long measurement time, even when working in the UHV pressure range. This effect is even amplified by the heat applied by the primary electron beam and the high oxygen affinity of Al. On the other hand, MnO is also being reduced by the electron beam and loses O after long measurement. This effect was tested and proven by taking Auger spectra from a MnO region after having been sputtered a second time (not shown here).

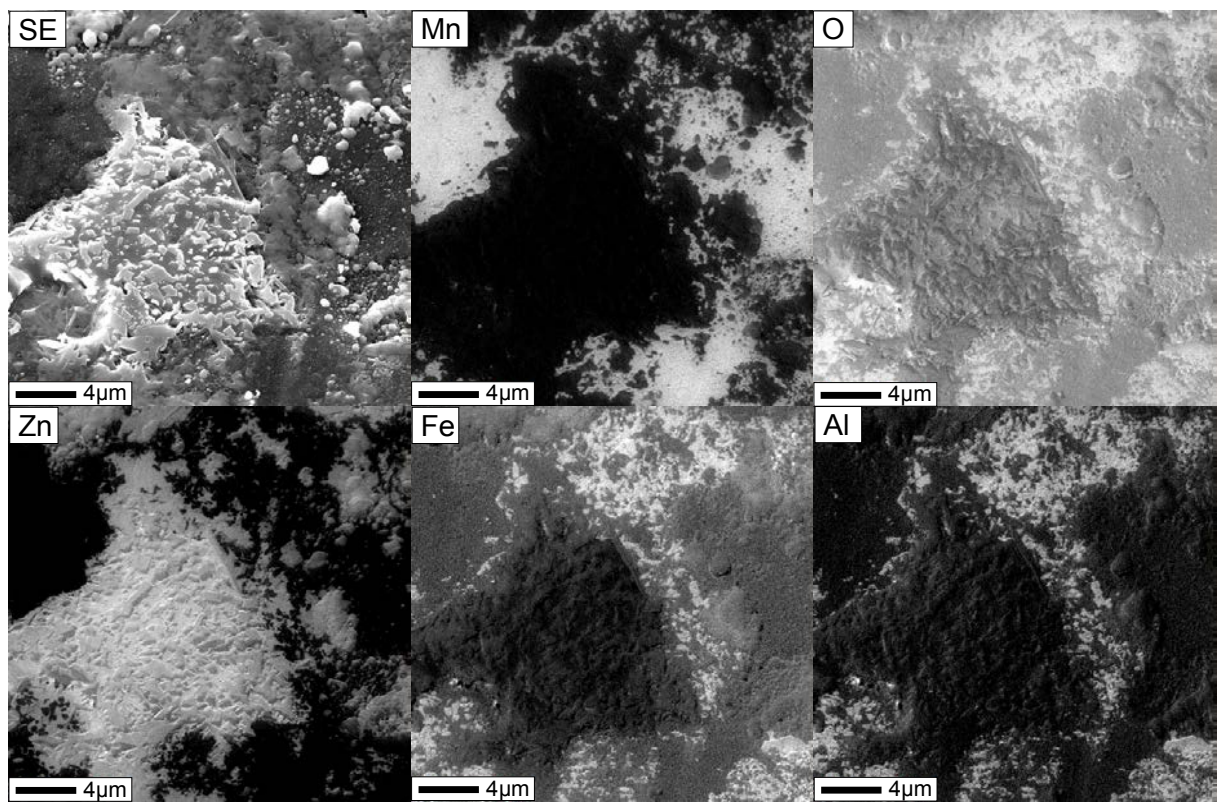

**Fig. 10:** SE overview image and AES mappings of the sputter cleaned surface in a region between Zn droplets.

## *Evaluation of the AES mappings from the sputter cleaned surface*

In order to identify all the different phases in the mappings shown in Figure 10 a special kind of presentation style is again needed. In this context the Al and O mappings were skipped since the Al signal gives no additional information and the O signal is, as stated above, completely misleading. However, the secondary electron signal does not only contain topographic information but also brightness (*e.g.* influenced by the work function) which can be used to identify oxides. In order to keep this information for the Zn, Fe, and Mn signals the data was processed according the rule I = (P-B). By that way the background signal is not lost, with the result shown in Figure 11.

For a proper identification and distinction between different phases AES spectra were taken at selective points to determine concentration differences. Concentrations taken from Spectra 1a and 1b in Figure 11 (represented by the blue colour) lead to the homogeneous MnO layer on the surface. The red colour and the spectra 2a and 2b are connected with the pure Zn metal phase. Spectra 3a and 3b show the $FeZn_{13}$ ζ phase in light brown. Spectra 4a and 4b mark crystals from the $Fe_2Al_5$ inhibition layer in green. Spectra 5a and 5b are taken on the bright round oxide structures which contain a mixture of approximately 20 at.% Zn, 9 at.% Fe, 21 at.% Mn, and 50 at.% O. The dark brown region from where the spectra 6a, 6b, and 6c were taken is a ZnFe phase, being not a ζ phase due to a higher Fe concentration of 20 at.%. Finally, the yellow area with spectra 7a, 7b, and 7c marks an oxidic region with an average concentration of 39 at.% Zn, 15 at.% Fe, 5 at.% Mn, and 41 at.% O.

Whether all these phases contain oxygen can not be directly deduced from the taken spectra. In order to confirm the findings an additional sputter cycle was performed and the spectra were again taken (not shown). As a result the oxygen concentration significantly decreased in the regions with spectra 2, 3, 4, and 6. In the region of spectrum 1, which is the MnO phase, the concentration increased slightly. Region 6 additionally showed also an increase of the Al signal. From these observations it can therefore be deduced that the brown areas are very thin and covering an underlying green $Fe_2Al_5$ inhibition layer phase.

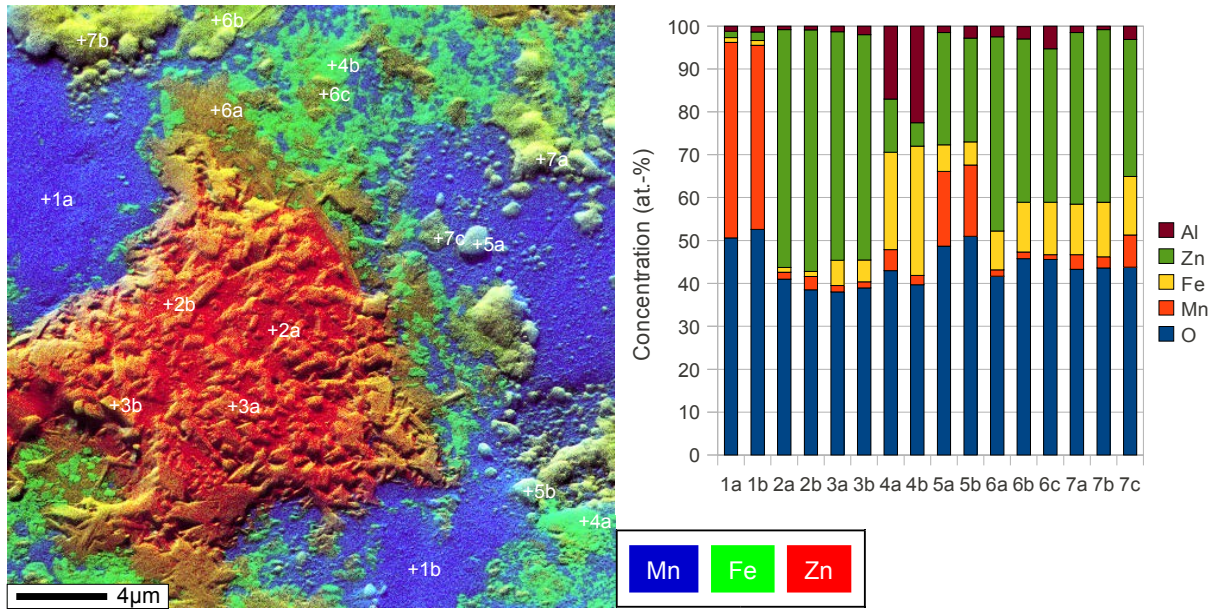

**Fig. 11:** The image on the left shows a combination of the intensity of the background signal, keeping the topography and work function signal, and the elemental information of Mn, Fe, and Zn, shown in blue, green, and red. On the right side an elemental quantification of spectra taken from different points in the mapping is shown.

## Discussion

In general, the first observation with respect to the nano TWIP steel sample is that the surface is not covered completely by zinc after the hot dip galvanising process, with the EDX results of Figure 1 revealing the reason for this behaviour. Wide areas covered by a continuous 100 – 200 nm thick MnO layer due to the high amount of Mn in the steel, as also indicated in literature, *e.g.* by Cho et al [11]. At low dew points during annealing prior the galvanization process it is expected that MnO forms a layer on top of the steel. From the Ellingham diagram [16] it is known that the formation of MnO from Mn and $O_2$ is energetic preferable to the formation of FeO. At high dew points during annealing also the Fe is oxidized on the surface and the segregation of Mn is stopped. The MnO distribution is in that case globular surrounded by Fe in the first few micrometres of the surface.

There are nevertheless areas on the surface which are covered by Zn in shape of extended droplets, with thicknesses of more than 20 μm. The surface of these Zn droplets itself is covered by a 5 nm thick homogeneous $Al_2O_3$ layer, being also in agreement with literature [17, 18]. Below these Zn droplets a $Fe_2Al_5$ inhibition layer is present, which not only protects the Zn from forming ZnFe mixed phases but also allows the successful bonding of the Zn to the surface.

The main difference between those two kinds of areas becomes clear when studying in detail the FIB cross preparation mappings: there are Fe platelets on the surface where the contact to the underlying bulk iron is interrupted, as seen in Figure 8d. The Zn even fills the gaps between these platelets and the bulk. Furthermore, in these small isolated platelets, there is not enough Mn in order to form a homogeneous MnO layer. In addition, when Fe is available at the surface then $Fe_2Al_5$ crystals can be formed and Zn can adhere.

Khondker et al. [19] introduced a model of an aluminothermic reduction of manganese oxides by Al in the Zn bath. As a hint for this model, Kawano et al. [20] found $Al_2O_3$ on top of MnO after contact with liquid Zn containing 0.2 at.% Al. Such a $Al_2O_3$ compound was now found in our work on the whole surface of the hot dip galvanised steel, however not after sputtering the surface with argon ions, as presented in Figure 11. Due to this model also the formation of the $FeZn_{13}$ ζ-phase is a hint for the aluminothermic reduction: Al is hindered to form $Fe_2Al_5$ crystals since it is present in oxidic form after the contact with MnO. When Fe is now coming in touch with Zn, these elements form the brittle intermetallic phase. In contrast to Kavitha et al. [12] no additional hint could be found in this work for the existence of an aluminothermic reduction, except the ζ-phase crystals, as depicted in Figure 9. As to $Al_2O_3$, this compound was found after sputtering on the surface as well as in some areas on top of a Zn droplet, see right upper corner of Figure 4, but not near the boundary between Zn and MnO. Additionally, on the surface mapping after ion sputtering (Figure 11) some mixed oxides containing Zn, Fe, and Mn, but no Al, could be identified.

A final observation in this work is the presence of Zn everywhere in the top nanometres on the surface, as shown in Figure 3 and 4. The presence of Zn could even be verified directly on top of the MnO regions. Consequently, within further experiments it needs to be verified if the Zn reaches the surface during being in contact with the bath or afterwards while being pulled out from the bath and being exposed to a Zn vapour phase.

## Conclusions

In this work the surface of a second generation advanced high strength steel AHSS, a so called nano TWIP, was analysed after hot dip galvanisation in a simulator by means of EDX, FIB and AES. Due to the low dew point during the annealing prior to the immersion into the zinc bath mayor parts of the surface of the twip steel are covered by a continuous 100-200 nm thick layer of MnO which prevents the growth of $Fe_2Al_5$ inhibition layer crystals and so the development of an uninterrupted zinc coating. The first nanometres of the sample surface are covered by an $Al_2O_3$ layer and Zn on top of Zn droplets as well as on the MnO layer, but no Fe could be found directly on the surface. In the MnO areas FeZn mixed crystals could be observed and identified as $FeZn_{13}$ ζ-phase. There was not found much evidence for an aluminothermic reduction of MnO except the ζ-phase. Moreover larger oxides near the surface did not contain Al but all other involved metals such as Fe, Mn, and Zn.

For the first time high quality chemical mappings in the nanometre range were made in this work on high manganese steel after hot dip galvanizing, not only on the surface but also in the surface near region by FIB cross sections and subsequent AES imaging. In combination with EDX with different electron beam acceleration voltages this analysis approach has led to a more complete understanding of the processes which take place in detail during the galvanizing procedure, helping in the future to promote the optimization of hot dip galvanised high manganese steel.

## Acknowledgements

The financial support by the Federal Ministry of Economy, Family and Youth and the National Foundation for Research, Technology and Development is gratefully acknowledged.